\documentclass[11pt]{article} 
\usepackage{amsmath,amsfonts,graphicx} 
\include{graphics}

\newcommand{\be}{\begin{align}}
\newcommand{\ee}{\end{align}}
\newcommand{\bear}{\begin{eqnarray}}
\newcommand{\eear}{\end{eqnarray}}

\newcommand{\tr}{\mathrm{Tr}}

\newcommand{\ba}{\begin{array}}
\newcommand{\ea}{\end{array}}

\newcommand{\nn}{\nonumber}
\newcommand{\diag}{\textrm{diag}}
\newcommand{\hs}{\textrm{Hubbard-Stratonovich}}
\newcommand{\ps}{\textrm{Pruisken-Sch\"afer}}
\newcommand{\s}{\ \! }

\begin{document}

\title{A conjecture on Hubbard-Stratonovich transformations for the Pruisken-Sch\"afer parameterisations
of real hyperbolic domains}
\date{}
\author{Yi Wei and Yan V Fyodorov\\
School of Mathematical Sciences, University of Nottingham, NG7 2RD,
UK} \maketitle

\abstract{Rigorous justification of the Hubbard-Stratonovich
transformation for the Pruisken-Sch\"afer type of
parameterisations of real hyperbolic $O(m,n)-$invariant domains
remains a challenging problem. We show that a naive choice of the
volume element invalidates the transformation, and put forward a
conjecture about the correct form which ensures the desired
structure. The conjecture is supported by complete analytic
solution of the problem for groups $O(1,1)$ and $O(2,1)$, and by a
method combining analytical calculations with a simple numerical
evaluation of a two-dimensional integral in the case of the group
$O(2,2)$.}

\section{Introduction and formulation of the conjecture}

For more than two decades, the nonlinear $\sigma$-model
methodology has been widely applied to studies of single electron
motions in disordered and chaotic mesoscopic
systems\cite{efe1,mirlin}. The method was pioneered by Wegner
\cite{weg} and further developed by Wegner and Sch\"afer
\cite{sw}, and Pruisken and Sch\"afer \cite{ps} in the framework
of the replica method used to reduce one-particle Hamiltonians
with microscopic disorder to a nonlinear $\sigma$-model. In the
early eighties, Efetov \cite{efe} introduced the supersymmetric
variant of the method which avoided the problematic replica trick
and directly led to the supermatrix version of the nonlinear
$\sigma$-model. Since then this latter nonlinear $\sigma$-model
has been also successfully applied to a variety of problems in the
framework of random matrix approach to chaotic scattering
\cite{vwz} \cite{fs}, Quantum Chromodynamics \cite{vw},
as well as a few other fields of physics.\\

A standard derivation of the nonlinear $\sigma$-models requires to
use at some point the so-called Hubbard-Stratonovich
transformation:
\be \label{eq:hs} C_n
e^{-\frac{1}{2}\tr\hat{A}^2}=\int \mathcal{D}\hat{R}\
e^{-\frac{1}{2}\tr\hat{R}^2 -i\tr \hat{R}\hat{A}}\,,
\end{align}
where $\hat{R}$ and $\hat{A}$ are $n\times n$ matrices and $C_n$
is a normalisation factor independent of the matrix $\hat{A}$.
When matrices $\hat{R}$ and $\hat{A}$ are, for example, complex
Hermitian, the volume element can be chosen as
$\mathcal{D}\hat{R}\propto\prod_{i\le
j}d\left[\mbox{Re}\,R_{ij}\right]\,d\left[\mbox{Im}\,R_{ij}\right]$,
and the above integral amounts to a product of standard Gaussian
integrals over independent degrees of freedom, the identity
(\ref{eq:hs}) following immediately. The same method works
obviously for the real symmetric matrices. On the other hand, in
these simple cases we also have a freedom to go to "polar"
coordinates in the standard way. For example, for the complex
Hermitian case \cite{mehta} \be \label{polar1}
\hat{R}=\hat{U}^{-1}\mathrm{diag}(p_{1},\ldots,p_{n})\hat{U},
\hspace{10mm}\mathcal{D}R\propto d\mu_H(U)dP\Delta^2[\hat{P}]\,,
\end{align}
where $\hat{U}\in \textrm U(n)$ is a unitary matrix of
eigenvectors, and $\hat{P}=\mathrm{diag}(p_{1},\ldots,p_{n})$ is
the real diagonal matrix of the associated eigenvalues of
$\hat{R}$, with $d\mu_H(U)$ being the corresponding invariant Haar
measure on the unitary group and $\Delta[\hat
P]=\prod_{i<j}(p_j-p_j)$ standing for the Vandermonde determinant
factor. Similarly, for the real symmetric matrices \be
\label{polar2} \hat{R}=\hat{O}^{-1}\hat{P}\hat{O},
\hspace{10mm}\mathcal{D}R\propto d\mu_H(O)dP|\Delta[\hat{P}]|\,,
\end{align}
with $\hat{O}\in \textrm O(n)$ being an orthogonal matrix.\\

In the problems of interest in electronic transport and random
matrix theory the structure of the matrices $\hat{R}$ and
$\hat{A}$ is however restricted by the underlying symmetries of
the system, and is rather non-trivial, see \cite{zirn-rev1} for a
review. For the simplest choice of the disordered Hamiltonian
corresponding to a system with broken time-reversal symmetry, one
of the legitimate choices of the integration domain for $R$ is due
to Sch\"afer and Wegner\cite{sw}: \be \label{eq:}\hat{R}=\lambda
\hat{T}\hat{T}^\dagger +i\hat{P},
\end{align}
where the matrices $\hat{T}$ must be chosen in the pseudounitary
group: $\hat{T}\in \textrm U(n_1,n_2)$. The matrices $\hat{P}$ are
Hermitian block-diagonal:
$\hat{P}=\mathrm{diag}(\hat{P}_{n_1},\hat{P}_{n_2})=\hat{P}^\dagger$,
and $\lambda>0$ is an arbitrary positive number. For Hamiltonians
respecting time-reversal symmetry the integration domain $\hat{R}$
is essentially of the same form, but with matrices $\hat{P}$ real
symmetric block-diagonal and the matrices $\hat{T}$ taken as
elements of the real pseudoorthogonal group: $\hat{T}\in \textrm
O(n_1,n_2)$.\\

Although the Sch\"afer-Wegner parameterisation of the integration
manifold is correct, an accurate verification of the main formula
Eq.(\ref{eq:hs}) is not at all trivial, and was provided only
recently\cite{zirn-rev1}. Actually, this type of parametrization
has never been widely used in the physical literature. Instead, an
alternative parameterisation due to Pruisken and Sch\"afer
\cite{ps} has been assumed, tacitly or explicitly, in the vast
majority of applications: \be
\label{eq:ps}\hat{R}=\hat{T}^{-1}\hat{P}\hat{T},
\hspace{10mm}\mathcal{D}R=d\mu_H(T)dP_1dP_2\Delta^2[\hat{P}].
\end{align}
Here we assumed the case of broken time-reversal symmetry,
$\hat{T}\in \textrm U(n_1,n_2)$ and
$\hat{P}=\mathrm{diag}(\hat{P}_{n_1},\hat{P}_{n_2})$, with
$\hat{P}_{n_1}$ and $\hat{P}_{n_2}$ being real diagonal,
$d\mu_H(T)$ being the invariant Haar measure on the pseudounitary
group and $\Delta[\hat P]=\prod_{i<j}(p_j-p_j)$ is the Vandermonde
determinant factor. Apparently, this parametrization is a complete
analogue of that in the formula (\ref{polar1}), specified for the
pseudo-unitary symmetry.\\

Similarly, one expects that a natural analogue of (\ref{polar2})
for the preserved time-reversal Hamiltonians and emerging
real-hyperbolic domain should be \be
\label{eq:ps1}\hat{R}=\hat{T}^{-1}\hat{P}\hat{T},
\hspace{10mm}\mathcal{D}R=d\mu_H(T)dP_1dP_2|\Delta[\hat{P}]|,
\end{align}
where this time $\hat{T}\in \textrm O(n_1,n_2)$ is the
corresponding pseudo-orthogonal matrices.\\

To the best of our knowledge, the validity of the
Hubbard-Stratonovich transformation with the Pruisken-Sch\"{a}fer
choice of the integration domain has not been carefully checked,
but rather taken for granted. In fact, the simplest version of the
"deformation of contour" argument used to verify the
transformation for the Sch\"afer-Wegner domain fails for the
Pruisken-Sch\"afer choice \cite{zirn-rev1}, and this raised
legitimate doubts on its validity in general, see also
\cite{zirn-sss}.\\

Given the widespread use of the Pruisken-Sch\"afer
parametrisation, as well as known technical advantages of working
with it in some microscopic models, the situation clearly calls
for further analysis. To this end, a rigorous proof of the
validity of the Hubbard-Stratonovich transformation for the
general pseudounitary Pruisken-Sch\"afer domain (\ref{eq:ps}) was
given for the first time by one of the authors \cite{yf1}. In the
same paper a variant of the Hubbard-Stratonovich transformation
for disordered systems with an additional chiral symmetry was also provided. \\

On the other hand, the problem of verifying Hubbard-Stratonovich
transformation for the general real pseudoorthogonal
Pruisken-Sch\"afer domain (\ref{eq:ps1}) turned out to be much
more challenging due to serious technical difficulties to be
discussed later on in the text of the paper. Only the simplest,
yet non-trivial case $\textrm O(1,1)$ was managed successfully in
\cite{yf1}, and we summarize the results of that study below. The
integration domain on the right hand side of Eq.~\eqref{eq:hs} is
given explicitly by
\be \hat{R}=\hat{T}^{-1}\hat{P}\hat{T}\, ,
\end{align}
where
\be
\hat{T}=\left(\begin{array}{cc}\cosh\theta & \sinh\theta\\
\sinh\theta &\cosh\theta
\end{array}\right)\in\frac{\textrm O(1,1)}{\textrm O(1)\times \textrm O(1)},
\ \ \textrm{and}\ \ \hat{P}=\textrm{diag}(p_1,p_2)\,.
\end{align}
The
matrices $\hat{A}$ in Eq.~\eqref{eq:hs} has the following form
\be\hat{A}=\left(\begin{array}{cc}a_1&-a\\ a&-a_2
\end{array}\right), \ \ {\mathrm{with}}\ \ a_1>0,\ a_2>0, \
|a|<\sqrt{a_1a_2}\ .
\end{align}
As has been shown in \cite{yf1} the desirable form (\ref{eq:hs})
of the $\hs$ transformation is only possible after one makes the
following choice of volume element on the integration manifold \be
d\hat{R}=(p_1-p_2)dp_1\ dp_2\ d\theta\,,
\end{align}
 whereas the would-be "natural" choice of the non-negative volume element
$$d\hat{R}=|p_1-p_2|dp_1\ dp_2\ d\theta\,,$$  as in (\ref{eq:ps1}),
can not yield a Gaussian function in the left-hand side of (\ref{eq:hs}). \\

In the present paper we continue that study by considering two
more specific cases - $O(2,1)$ and $O(2,2)$, and investigating in
detail the validity of the $\hs$ transformation for the
corresponding real hyperbolic domains. Note that for practical
needs of the theory of disordered systems $O(2,2)$ is the most
important case related, in the supersymmetric version, to the
basic object of the theory, the so-called two-point correlation
function of resolvents of the random Schroedinger operator, see
e.g. \cite{efe1,zirn-rev1}.\\

In both $O(2,1)$ and $O(2,2)$ cases we are able to show that the
naive choice of the measure Eq.(\ref{eq:ps1}) is never possible,
but the $\hs$ transformation (\ref{eq:hs}) can be saved provided
we make a suitable alternative choice of $D\hat{P}$.  These
examples naturally suggest to put forward the following conjecture
on the correct form of the $\hs$ transformation on a general
$\textrm O(m,n)-$invariant Pruisken-Sch\"{a}fer domain. Define \be
\label{eq:psconja} \hat{R}=\hat{T}^{-1}\hat{P}\hat{T},\quad
\hat{P}&=\diag(\hat{P}_1,\hat{P}_2)=\diag\s(p_{11}, \dots,
p_{1m},p_{21},\dots, p_{2n})
\end{align}
and the volume element \be \label{eq:psconjb}
\mathcal{D}R=d\mu_H(T)\,\mathcal{D}\hat{P}, \quad
\mathcal{D}\hat{P}&=|\Delta[\hat{P}_1]|\cdot|\Delta[\hat{P}_2]|\prod_{i=1}^m\prod_{j=1}^n(p_{1i}-p_{2j})\,
,
\end{align}
where $|\Delta[\hat{P}]|$ is the absolute value of the Vandermonde
determinant,  and $d\mu_H(\hat{T})$ stands for the invariant
measure on $O(m,n)$. Further assume that the the real matrix
$\hat{A}$ is of the form $\hat{A}=\hat{A}_+\hat{L}$, where
$\hat{A}_+$ is positive definite and $\hat{L}$ is the signature
matrix $\hat{L}$ appearing in the definition of the
pseudoorthogonal group $O(m,n)$\footnote{Such matrices can always
be brought to a real diagonal form by $O(m,n)$ rotations, see e.g.
Appendix B of the paper \cite{fiz}.}. Then the $\hs$
transformation over the $\ps$ type of real hyperbolic domain is
given by \be \label{eq:psconjc} \int \mathcal{D}\hat{R}\
e^{-\frac{1}{2}\tr\hat{R}^2 -i\tr
\hat{R}\hat{A}}&=\int_{-\infty}^{\infty} \mathcal{D}\hat{P}\
e^{-\frac{1}{2}\left[\sum\limits_{i=1}^{m}p_{1i}^2+\sum\limits_{j=1}^{n}p_{2j}^2\right]}
\int_{\mathrm{O}(m,n)}d\mu_H(\hat{T})
e^{-i\tr\hat{T}^{-1}\hat{P}\hat{T}\hat{A}}
\end{align}
\[={\mathrm{const.}}\ e^{-\frac{1}{2}\tr \hat{A}^2}\s .\]

The formula Eq.~\eqref{eq:psconjc} is the central message of our
work. The crucial difference of the choice (\ref{eq:psconjb}) from
the naive choice of the measure (\ref{eq:ps1}) is the absence of
modulus for the factors
$\prod_{i=1}^m\prod_{j=1}^n(p_{1i}-p_{2j})$. This forces the
volume element to change sign inside the integration domain, in
contrast to the conventional measures (densities) which are always
positive as in e.g. Eq.(\ref{polar2}). Such feature does not
however in any way invalidate our Hubbard-Stratonovich formula,
which should be interpreted as follows. The actual sign of
$\mathcal{D}\hat{R}$ is determined by the inequalities between
$p_{1}$'s and $p_{2}$'s. An ordered sequence of the $p_{1}$'s and
$p_{2}$'s thus defines a sub-domain of $\hat{R}$ on which the sign
of $\mathcal{D}\hat{R}$ is fixed. Without loss of generality, we
can assume $p_{11}>p_{12}>\cdots>p_{1m}$ and
$p_{21}>p_{22}>\cdots>p_{2n}\,$. Then it is clear that the domain
of integration in $\hat{R}$ is a union of altogether $(m+n)!/m!n!$
such disjoint sub-domains. Labelling a particular choice of the
sub-domain of this sort by $D_\sigma$ and defining
$\mathrm{sgn}(\sigma)$ to be the sign of the volume element
$\mathcal{D}\hat{R}$ on $D_\sigma\,$, the left-hand side of the
integration formula we discuss is given by \be \int
\mathcal{D}\hat{R}\ e^{-\frac{1}{2}\tr\hat{R}^2 -i\tr
\hat{R}\hat{A}}=\sum_{\sigma}\mathrm{sgn}(\sigma)\int_{D_\sigma}
|\mathcal{D}\hat{R}|\ e^{-\frac{1}{2}\tr\hat{R}^2 -i\tr
\hat{R}\hat{A}}\;.
\end{align}

Interpreting our formula in this way, we always integrate over
each sub-domain $D_\sigma$ with the well-defined positive measures
$|\mathcal{D}\hat{R}|$, but the l.h.s. of Eq.~\eqref{eq:psconjc}
is given by an alternating sum of integrals on the disjoint
sub-domains of $\hat{R}$.  We believe this coordinated change of
sign is absolutely necessary to ensure the Gaussian form of the
result of the integration, the conviction being based on the
example of \cite{yf1} and the results of the current paper.\\

We consider verification of this conjecture, as well as the
discovery of a general mechanism which ensures its validity to be
a challenging problem reserved for a future research \footnote{ A
method of proving the validity of the above conjecture in the
general case $O(m,n)$ has recently been proposed by M. R.
Zirnbauer and the present authors, and will be published
elsewhere\cite{rigor}.}.

\section{Verification of the conjecture for O(2,1) case}
In this section, we consider the $\ps$ type of parameterisation of
integration domain Eq.~(\ref{eq:psconja}) with $\hat{T}$ being an
element of the real pseudoorthogonal group $\textrm{O}(2,1)$. The
real matrix $\hat{A}$ in Eq.~(\ref{eq:psconjc}) is assumed to be
of the form $\hat{A}=\hat{A}_+\hat{L}$, where $\hat{A}_+$ is
positive definite and $\hat{L}$ is the signature matrix
$\hat{L}=\diag(1,1,-1)$. As mentioned above, such matrices
$\hat{A}$ can be always diagonalised as
$\hat{A}=\hat{T}^{-1}\Lambda\hat{T}$, with
$\hat{T}\in\mathrm{O}(2,1)$ and $\Lambda$ is a real diagonal
matrix. By exploiting the invariance of the Haar measure we can
safely choose $\hat{A}$ to be diagonal, as this choice obviously
does not change the result of
the integration.\\

Implementing the $\ps$ parametrisation , the integral on the right
hand side of Eq.~(\ref{eq:psconjc}) is of the form of
\be
\label{eq:break} I_{HS}^{\mathrm{O}(2,1)}=\int \mathcal{D}\hat{R}\
e^{-\frac{1}{2}\tr\hat{R}^2 -i\tr
\hat{R}\hat{A}}=\int_{-\infty}^{\infty} \mathcal{D}\hat{P}
e^{-\frac{1}{2}\sum\limits_{i=1}^{3}
p_i^2}\int_{\mathrm{O}(2,1)}d\mu(\hat{T})
e^{-i\tr\hat{T}^{-1}\hat{P}\hat{T}\hat{A}}\, ,
\end{align}
 where $\hat{P}=\diag(p_1,p_2,p_3)$ and $d\mu(\hat{T})$ is the invariant Haar measure on $\mathrm{O}(2,1)$.
 The crucial point is that we have to choose the volume element
$\mathcal{D}\hat{P}$ to be, cf. Eq.~(\ref{eq:psconjb}),
\be
\label{eq:dP}\mathcal{D}\hat{P}=|p_1-p_2|(p_1-p_3)(p_2-p_3)dp_1\s
dp_2\s dp_3.
\end{align}
 We are going to demonstrate that it is only this choice that validates the $\hs$
 transformation for our choice of the hyperbolic domain.\\

Note that the integral over the pseudoorthogonal group
$\mathrm{O}(2,1)$ on the right hand side of Eq.~\eqref{eq:break}
is of the type of the Harish-Chandra-Itzykson-Zuber integral.
Although integrals of this type have been known long ago for
unitary groups \cite{iz} and extended more recently to
pseudounitary groups \cite{fes}, their analogues for
(pseudo)orthogonal groups, which is relevant here, remains largely
an open problem in mathematical physics, although a few
interesting insights were obtained very recently
\cite{GK,BH}.

\subsection{Particular example of the O(2,1) $\hs$ transformation}
To elucidate main points of the calculation we first consider a
special choice of the (diagonal) matrix $\hat{A}$, that is
\be\label{eq:simp} \hat{A}=\diag(x,x,z) \implies
e^{-\frac{1}{2}\tr \hat{A}^2}=e^{-\frac{1}{2}(2x^2+z)}\, .
\end{align}
Since $\hat{A}\hat{L}=\diag(x,x,-z)>0$ according to our
assumption, we have to require $x>0>z$.\\

The calculations will be simpler as such $\hat{A}$ effectively
replaces the
integration over the whole group $\mathrm{O}(2,1)$ with one over the non-compact
Riemannian symmetric space
\be \frac{\mathrm{O}(2,1)}{\mathrm{O}(2)\times\mathrm{O}(1)}\cong\frac{\mathrm{SO}
(2,1)}{\mathrm{S}[\mathrm{O}(2)\times\mathrm{O}(1)]}\,.
\end{align}
Denote $d\mu(\hat{S})$ the $\mathrm{O}(2,1)$ invariant measure on the non-compact
Riemannian symmetric space $G/H$, with $G=\mathrm{O}(2,1)$ and
$H=\mathrm{O}(2)\times\mathrm{O}(1)$.
For our special choice of the matrix $\hat{A}$ we obviously have
\be \label{eq:coset}\int_{\mathrm{O}(2,1)}d\mu(\hat{T}) \
e^{-i\tr\hat{T}^{-1}\hat{P}\hat{T}\hat{A}}=\int_{G/H}d\mu(\hat{S})\
e^{-i\tr\hat{S}^{-1}\hat{P}\hat{S}\hat{A}}\,,
\end{align}
so that Eq.~\eqref{eq:break} assumes the following form \be
\label{eq:simple} \int \mathcal{D}\hat{R}\
e^{-\frac{1}{2}\tr\hat{R}^2 -i\tr
\hat{R}\hat{A}}=\int_{-\infty}^{\infty} \mathcal{D}\hat{P}\
e^{-\frac{1}{2}\sum\limits_{i=1}^{3}p_i^2}
\int_{G/H}d\mu(\hat{S})\ e^{-i\tr\hat{S}^{-1}\hat{P}\hat{S}\hat{A}}. \end{align}\\

To perform the integration over the coset space $G/H$ it is
convenient to parametrise $G/H$ with the projective coordinates
$(Z,Z^T)$. To this end, we introduce a $2\times1$ real matrix $Z$
as
\be\label{Z} Z=\left(\begin{array}{c} z_1 \\
z_2\end{array}\right)\ \ {\mathrm{with\,\, the\,\, constraint}}\ \
1-Z^TZ\ge0\,,
\end{align}
in terms of which the matrices $\hat{S}$ on $G/H$ are given by \be
\label{eq:s} \hat{S}=\left(\begin{array}{cc}
(1-ZZ^T)^{-\frac{1}{2}} & Z(1-Z^TZ)^{-\frac{1}{2}}
\\ Z^T(1-ZZ^T)^{-\frac{1}{2}} & (1-Z^TZ)^{-\frac{1}{2}} \end{array}\right)\, .
 \end{align}

It is direct to check that $\hat{S}^{-1}(Z,Z^T)=\hat{S}(-Z,-Z^T)$.
The invariant measure $d\mu(\hat{S})$ in projective coordinates
can be calculated in the standard way\cite{hua} and is given by
\be d\mu(\hat{S})=\frac{dZdZ^T}{(1-Z^TZ)^{\frac{3}{2}}}\,,
\end{align}
where $dZdZ^T=dz_1dz_2$ and the integration domain is as specified
in (\ref{Z}). Make the following change of variables \be
\left\{\begin{array}{l}z_1=r\cos\theta \\
z_2=r\sin\theta\end{array}\right., \ \ \ \ r\in[0,1]\
{\mathrm{and}} \ \theta\in[0,2\pi]. \ \end{align} The integration
on the right hand side of Eq.\eqref{eq:coset} can be written as
\be\label{eq:inter} \int_{0}^{1}
\frac{rdr}{(1-r^2)^{\frac{3}{2}}}\int_{0}^{2\pi}d\theta\
\exp\frac{i}{2}\bigg\{&\frac{r^2}{1-r^2}(x-z)(p_1-p_2)\cos2\theta+\frac{x-z}{1-r^2}(p_1+p_2-2p_3) \nn\\
&+\left[x(p_1+p_2+2p_3)+z(p_1+p_2)\right] \bigg\}.\end{align} The
integral over $\theta$ yields the standard  Bessel functions in
view of $ \int_{0}^\pi d\phi\ e^{i\beta\cos \phi}=\pi
J_0(\beta)\s$, and introducing a new variable
$t=\frac{r^2}{1-r^2}$, we rewrite (\ref{eq:inter}) as \be
\label{eq:int-GH} e^{i(x(p_1+p_2)+zp_3)}
\int_{0}^{\infty}\frac{dt}{\sqrt{1+t}}J_0\left[\frac{t}{2}(x-z)(p_1-p_2)\right]e^{\frac{it}{2}(x-z)(p_1+p_2-2p_3)}\,.
\end{align}
Now we need to substitute Eq.~\eqref{eq:int-GH} into the right
hand side of Eq.~\eqref{eq:simple} and to integrate over
$\hat{P}$, that is \be
I_{HS}^{\mathrm{O}(2,1)}=\int_{0}^{\infty}\frac{dt}{\sqrt{1+t}}\int_{-\infty}^{\infty}
\mathcal{D}\hat{P}\
&\exp\bigg\{-\frac{1}{2}\sum\limits_{i=1}^{3}p_i^2+i(x(p_1+p_2)+zp_3)\nn\\
&+\frac{it}{2}(x-z)(p_1+p_2-2p_3)\bigg\} J_0\left[\frac{t}{2}(x-z)(p_1-p_2)\right]\, .\label{eq:sub}
\end{align}
After a straightforward, but lengthy calculation we arrive at the
following result \be I_{HS}^{\mathrm{O}(2,1)}=
\frac{\sqrt{2}\pi}{32}F[(x-z)^2]e^{-\frac{1}{2}(2x^2+z^2)}\label{eq:pre-gauss-1}\,,\end{align}
where
\be F(a)=\int_{0}^{\infty}\frac{dt}{\sqrt{1+t}}
\exp\left(-\frac{1}{2}(t^2+t)a\right)\left[1-a(2t^2+3t+1)\right].
\end{align}
Note that the expression Eq.~\eqref{eq:pre-gauss-1} contains already the Gaussian
factor of precisely the form required by (\ref{eq:simp}). Unfortunately, that factor is multiplied with a
function $F[(x-z)^2]$ dependent on the
combination
$a=(x-z)^2$, the fact seemingly incompatible with the $\hs$
transformation. Miraculously enough, this factor is an
$a-$independent constant! To verify this, we define
$y=\sqrt{1+t}$, and carry out the integral explicitly:
\be\label{eq:o21}
F(a)&=\int_{0}^{\infty}\frac{dt}{\sqrt{1+t}}\exp\left(-\frac{1}{2}(t^2+t)a\right)
\left[1-a(2t^2+3t+1)\right]\nn\\ &=\int_{1}^{\infty}dy\
\exp\left(-\frac{a}{2}(y^4-y^2)\right)
\left[1-a(2y^4-y^2)\right]\nn\\&=1-\lim_{y\to\infty}y\exp(-\frac{ay^2(y^2-1)}{2})=1\,.\end{align}
At the last step, we used the fact that $a$ is strictly positive,
as the case $a=0$ should be excluded from the very beginning.
Indeed, $a=0$ implies $x=z$, contradicting to the original
requirement $x>0>z$.

\subsection{General calculation for O(2,1) case}

Now we are ready to present the complete proof of the $\hs$
transformation over O(2,1) domain. In the general case we have
$\hat{A}=\diag(x_1,x_2,z)=\hat{A}_1+\hat{A}_2$ where
$\hat{A}_1=\diag(x,x,z)$ is the part considered in the previous
example, and $\hat{A}_2=\diag(w,-w,0)$. Here we defined the
variables $x=(x_1+x_2)/2$, $w=(x_1-x_2)/2$. Our starting point is
again Eq.~\eqref{eq:break}, but we now have \be
\label{eq:complete} I_{HS}^{\mathrm{O}(2,1)}&=\int
\mathcal{D}\hat{R}\
e^{-\frac{1}{2}\tr\hat{R}^2 -i\tr \hat{R}\hat{A}}\nn\\
&=\int_{-\infty}^{\infty} \mathcal{D}\hat{P} e^{-\frac{1}{2}
\sum\limits_{i=1}^{3}p_i^2}\int_{G/H}e^{-i\tr\hat{S}^{-1}\hat{P}\hat{S}\,\hat{A}_1}d\mu(\hat{S})\int_{H}d\mu(\hat{H})
\
e^{-i\tr\hat{S}^{-1}\hat{P}\hat{S}[\hat{H}\hat{A}_2\hat{H}^{-1}]},\end{align}
where we assume $G=\mathrm{O}(2,1)$, $H=\rm O(2)\times O(1)$ and
$S=G/H$ as before.\\

The integration over $H$ goes effectively over the group
SO(2) and the corresponding matrices can be parametrized in
a standard way as $H=\left(\begin{array}{cc}\cos\phi & \sin\phi\\
-\sin\phi& \cos\phi\end{array}\right)$. Using the same parameters
for the coset matrices $\hat{S}$ as in the previous section, we
then find \be \tr\
\hat{S}^{-1}\hat{P}\hat{S}\hat{H}\hat{A}_2\hat{H}^{-1}=A\cos2\phi+B\sin2\phi\, ,\end{align}
where \be
A=&\frac{w}{4(1-r^2)}\bigg\{\big[(1+\sqrt{1-r^2})^2+2r^2\cos2\theta+
\cos4\theta(1-\sqrt{1-r^2})^2\big]p_1\nn\\
&+\big[2r^2\cos2\theta-(1+\sqrt{1-r^2})^2-
\cos4\theta(1-\sqrt{1-r^2})^2\big]p_2-4r^2\cos2\theta p_3 \bigg\}\nn\\
B=&\frac{-w}{4(1-r^2)}\bigg\{\big[2r^2\sin2\theta+\sin4\theta(1-\sqrt{1-r^2})^2\big] p_1+
\big[2r^2\sin2\theta-\sin4\theta(1-\sqrt{1-r^2})^2\big]p_2\nn\\ &-4r^2\cos2\theta p_3 \bigg\}\, .
\end{align}
The integration over $\phi$ is easily performed according to the
formula \be \label{bessel11}
J_0(\sqrt{A^2+B^2})=\frac{1}{\pi}\int_0^\pi d\phi
\exp\big(i\cos\phi A+i\sin\phi B\big),
\end{align}
so that
 \be\int_{H} d\mu(\hat{H})
e^{-i\tr\hat{S}^{-1}\hat{P}\hat{S}\hat{H}\hat{A}_2\hat{H}^{-1}}=J_0(\sqrt{A^2+B^2})\,.
\end{align}

This should be inserted into Eq.~\eqref{eq:complete}, and
remembering Eq.~\eqref{eq:inter}-\eqref{eq:sub}, we arrive at \be
I_{HS}^{\mathrm{O}(2,1)}=\int_{0}^{\infty}\frac{dt}{\sqrt{1+t}}&
\int_{-\infty}^{\infty} \mathcal{D}\hat{P}\
\int_{0}^{2\pi}d\theta\
\exp\bigg\{-\frac{1}{2}\sum\limits_{i=1}^{3}p_i^2+i(x(p_1+p_2)+zp_3)\nn\\
&+\frac{it}{2}(x-z)(p_1+p_2-2p_3)+\frac{it}{2}(x-z)(p_1-p_2)\cos\theta\bigg\}
J_0(\sqrt{A^2+B^2})\label{eq:inapp},\end{align} where again
$\mathcal{D}\hat{P}$ is given by Eq~\eqref{eq:dP}.

Note that
variable '$w$' responsible for the difference from the example
considered in the previous section enters the formula only via the
combination $\sqrt{A^2+B^2}$.  A way of evaluating the above
integral for $w\ne 0$ is to expand the Bessel function in Taylor
series with the n-th term proportional to $w^{2n}$, to integrate
each term separately, and then re-sum the series. A straightforward
implementation of this program is however not immediate, and
necessary steps of the proof are given in
App.\ref{app:completeo21} where it is shown that
\be\label{eq:o2,1} I_{HS}^{\mathrm{O}(2,1)}=\mathrm{const}\,
\exp\bigg[-x^2-w^2-\frac{z^2}{2}\bigg]=\mathrm{const}\,\exp\bigg[-\frac{1}{2}(x_1^2+x_2^2+z^2)\bigg]\, ,
\end{align}
in precise agreement with the structure required by the
Hubbard-Stratonovich transformation.\\

To summarize, we have demonstrated that for any
$\hat{A}=\hat{T}_0\s\diag(x_1,x_2,z)\s\hat{T}^{-1}_0$ and
$\hat{T}_0\in\mathrm{O}(2,1)$ holds the identity \be \int
\mathcal{D}\hat{R}\ e^{-\frac{1}{2}\tr\hat{R}^2 -i\tr
\hat{R}\hat{A}}=\mathrm{const}\,\ e^{-\frac{1}{2}\tr\hat{A}^2},
\end{align}
provided the volume element ${\cal D}P$ for the $\hat{P}$ integral
is
chosen in accordance with Eq.~\eqref{eq:dP}. \\

For the sake of comparison, one may try to repeat the above
calculation with the "naive" choice of measure
$D\hat{P}=|\Delta(\hat{P})|\prod_{i=1}^3dp_i$  instead of
Eq.~\eqref{eq:dP}. We show in App.~\ref{app:check1} that such a
choice invalidates the $\hs$ transformation. As another
comparison, we also provide similar calculations in
App.~\ref{app:o3} for the compact counterpart of this $\ps$ domain
corresponding to the group O$(3)$.

\section{Results for the O(2,2) case}
In this section, we carry out the detailed calculation for the
$\hs$ transformation over the O$(2,2)$ $\ps$ domain. As the
calculation turns out to be quite technically cumbersome, we
restrict ourselves with the simplest non-trivial choice
 $\hat{A}=\diag(x,x,z,z)$, with $x>0>z$. Consequently, the integration domain
$\hat{T}=\mathrm{O}(2,2)$ effectively reduces to the non-compact
Riemannian symmetric space (coset space) \be \rm
\frac{O(2,2)}{O(2)\times O(2)}\cong\rm \frac{SO(2,2)}{S[O(2)\times
O(2)]}\s.\end{align} Parameterisation of $G/H$, where
$G=\mathrm{SO}(2,2)$ and $H=\mathrm{S[O(2)\times O(2)]}$, with the
projective coordinates $Z$ and $Z^T$ is again in the form of
Eq.~\eqref{eq:s} with $Z$ and $Z^T$ being real $2\times2$ matrices
chosen in a way ensuring that the matrix $1-Z^TZ$ is positive
definite: \be Z=\left(\begin{array}{cc} z_1&z_2 \\
z_3&z_4\end{array}\right)\ \ {\mathrm{with}}\ \ 1-Z^TZ\ge 0\,.
\end{align}

We aim to prove the validity of the $\hs$ transformation with the
$\ps$ parameterisation Eq.~\eqref{eq:ps}, where
$T\in\mathrm{O}(2,2)$ and $\hat{P}=\diag(p_1,p_2,p_3,p_4)$. To
this end, we need to demonstrate that the following integral \be
I_{HS}^{\mathrm{O}(2,2)}=\int \mathcal{D}\hat{R}\
e^{-\frac{1}{2}\tr\hat{R}^2 -i\tr
\hat{R}\hat{A}}&=\int_{-\infty}^{\infty} \mathcal{D}\hat{P}\
e^{-\frac{1}{2}\sum\limits_{i=1}^{4}p_i^2}\int_{\mathrm{O}(2,2)}
d\mu(\hat{T}) e^{-i\tr\hat{T}^{-1}\hat{P}\hat{T}\hat{A}} \nn\\
&=\int_{-\infty}^{\infty} \mathcal{D}\hat{P}\ e^{-\frac{1}{2}
\sum\limits_{i=1}^{4}p_i^2} \int_{G/H}d\mu(\hat{S})\
e^{-i\tr\hat{S}^{-1}\hat{P}\hat{S}\hat{A}}
\label{eq:o22}\end{align}
 is, up to a constant factor, a product of Gaussian factors.
The invariant measure $d\mu(\hat{S})$ here is calculated in the
standard way and is equal to \cite{hua}
\be d\mu(\hat{S})=\frac{dZdZ^T}{\det(1-Z^TZ)^{2}},\end{align}where $dZdZ^T=dz_1dz_2dz_3dz_4$. \\

To carry out the integration over the coset space we introduce the
polar coordinates parametrization for real matrices $Z$. This
amounts to diagonalizing $Z$ by two orthogonal rotations as
\be\label{eq:polar}
Z=O_1\left(\begin{array}{cc}r&0\\0&s\end{array}\right)O_2,\ \
{\rm{where}}\ \ r,s\in(-\infty,\infty),\ \
O_1,O_2\in\mathrm{SO}(2)\, .
\end{align}
A standard calculation (App.~\ref{app:jacobi}) shows that the Jacobian
induced by changing variables from $Z,Z^T$ to the polar
coordinates is simply $|r^2-s^2|$. We have accordingly \be
dZdZ^T=|r^2-s^2|\s dr\s ds\s d\mu(O_1)d\mu(O_2),
\end{align}where $d\mu(O_1)$ and $d\mu(O_2)$ are the invariant Haar measure of SO$(2)$.
Using the polar coordinates the integral over coset space takes
the form \be& \int_{G/H}d\mu(\hat{S})\
e^{-i\tr\hat{S}^{-1}\hat{P}\hat{S}\hat{A}}\nn\\=&\int
D(r,s)\int_{\mathrm{SO}(2)}d\mu(O_1)\exp\left\{ i\tr \left[
O_1\left(\begin{array}{cc}\frac{x-zr^2}{1-r^2}&0\\0&\frac{x-zs^2}{1-s^2}
\end{array}\right)O_1^{-1}\left(\begin{array}{cc}p_1&0\\0&p_2\end{array}
\right)\right]\right\}\nn\\
&\hspace{16mm}\int_{\mathrm{SO}(2)}d\mu(O_2)\exp\left\{ i\tr \left[
O_2^{-1}\left(\begin{array}{cc}\frac{z-xr^2}{1-r^2}&0\\0&\frac{z-xs^2}{1-s^2}
\end{array}\right)O_2\left(\begin{array}{cc}p_3&0\\0&p_4\end{array}
\right)\right]\right\}\, ,
\label{eq:int-coset}
\end{align}
where we denoted $D(r,s)=|r^2-s^2|drds/(1-r^2)^2(1-s^2)^2$.\\

The two integrals over O$(2)$ group manifold in
Eq.~\eqref{eq:int-coset} are easily carried out using the formula
\be &\int_{\mathrm{SO}(2)}d\mu(O) \exp\left\{i\tr \
 O\left(\begin{array}{cc}a_1&0\\0&a_2\end{array}\right)O^{-1}
 \left(\begin{array}{cc}b_1&0\\0&b_2\end{array}\right)\right\}
\nn\\=&\ \exp\left[\frac{i}{2}(a_1+a_2)(b_1+b_2)\right]J_0
\left[\frac{1}{2}(a_1-a_2)(b_1-b_2)\right].\end{align} Introducing
at the next step the variables $u=\frac{1}{1-r^2}$ and
$v=\frac{1}{1-s^2}$, we rewrite the resulting integral in
Eq.~\eqref{eq:int-coset} as \be \label{eq:coset2}
e^{ix(p_3+p_4)+iz(p_1+p_2)}& \int_1^\infty
\frac{|u-v|dudv}{\sqrt{u(u-1)}\sqrt{v(v-1)}}\exp\left\{\frac{i}{2}(x-z)(p_1+p_2-p_3-p_4)(u+v)\right\}\nn\\
&J_0\left[\frac{1}{2}(x-z)(p_1-p_2)(u-v)\right]J_0\left[\frac{1}{2}(x-z)(p_3-p_4)(u-v)\right].
\end{align}
Now we have to perform the integration over variables in
$\hat{P}$. As in the previous section, the crucial point is to
choose the volume element $D\hat{P}$ in accordance with our main
conjecture, that is \be
D\hat{P}=|p_1-p_2|(p_1-p_3)(p_1-p_4)(p_2-p_3)(p_2-p_4)|p_3-p_4|\prod_{i=1}^4
dp_i\s .\label{eq:dP2}
\end{align}
The remaining steps are lengthy but straightforward. After a few
variable changes we arrive at

\be I_{HS}^{\mathrm{SO}(2,2)}=\frac{\pi}{128}{\cal
F}\left[a\right]\exp\left[-x^2-z^2\right]\label{eq:final}\,,\end{align}
with $a\equiv x-z$ and the function ${\cal F}[a]$ given in terms of
a double integral as \begin{eqnarray} {\cal F}[a] =\int_{1}^\infty dt \,
e^{ -\frac{a^2(t^2-1)}{4}} \int_{0}^{(t-1)^2}
\frac{\frac{1}{4}a^4t^2(t^2-v)-a^2t^2+1}{\sqrt{[(t+1)^2-v][(t-1)^2-v]}}\,
e^{-\frac{a^2v}{4}}\,dv\,\,.
\end{eqnarray}
Integrating over $v$ and defining a new variable $x=\frac{t+1}{2}$, we get
\be {\cal F}[a] =\frac{\pi}{128}\int_{1}^\infty\! dx&
\ e^{-a^2(x^2-x)} \frac{x-1}{x}\bigg\{\big[a^2(2x-1)^2-2\big]^2\Phi_1\big[1,\frac{1}{2},\frac{3}{2},\big(\frac{x-1}{x}\big)^2,-a^2(x-1)^2\big]
\nn\\ & -\frac{8}{3}a^4(x-1)^2(2x-1)^2\Phi_1\big[2,\frac{1}{2},\frac{5}{2},\big(\frac{x-1}{x}\big)^2,-a^2(x-1)^2\big]\bigg\}\ ,
\end{align}where $\Phi_1$ is the degenerate hypergeometric series of two variables defined as \cite{grad} \be \Phi_1[\alpha,\beta,\gamma, x,y]\ =\sum_{m,n=0}^\infty\frac{(\alpha)_{m+n}(\beta)_m}{(\gamma)_{m+n}\ m!n!}x^my^n\ .\end{align}

\begin{figure}
\centering
\includegraphics[width=5cm, height=4cm]{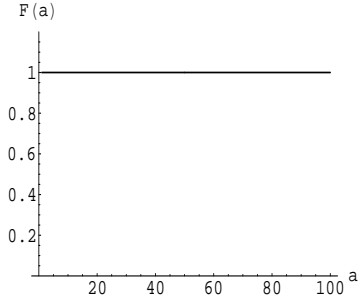}
\caption{Function $F(a)=1$ is a constant which does not depend on
$a$.} \label{f(a)}
\end{figure}

From Eq.~\eqref{eq:final} we see that only if the factor ${\cal
F}[a]$ is independent of its argument $a\equiv x-z$ the
whole expression $I_{HS}^{\mathrm{SO}(2,2)}$ can be in the desired
Gaussian form. It needs only a few lines of Maple or Mathematica
code to check numerically that actually ${\cal F}[a]\equiv 1$, see
Fig.\ref{f(a)}.

\hspace{50mm}

Unfortunately, we were not able to find a way of verifying this
miraculous identity analytically, as we managed to do in the
previous case of O$(2,1)$ integral. Nevertheless, we do not
think the numerical data leave any doubt in the validity of our claim. \\

In conclusion, the above calculation shows that for
$\hat{A}=\hat{T}_0\ \! \diag(x,x,z,z)\ \! \hat{T}^{-1}_0$ and $\hat{T}_0\in\mathrm{O}(2,2)$,
 \be \int \mathcal{D}\hat{R}\ e^{-\frac{1}{2}\tr\hat{R}^2
-i\tr \hat{R}\hat{A}}=\mathrm{const}\,\
e^{-\frac{1}{2}\tr\hat{A}^2},
\end{align}
provided measure for $\hat{P}$ integral is chosen to be Eq.~\eqref{eq:dP2}. \\

It is again interesting to check what will be the result if we
choose $d\hat{P}=|\Delta(\hat{P})|\prod_{i=1}^4dp_i$ instead of
Eq.~\eqref{eq:dP2}. It is shown in App.~\ref{app:check2} that this
choice will make the $\hs$ identity invalid.

\section*{Acknowledgements}
The research was supported by EPSRC grant EP/C515056/1"Random
Matrices and Polynomials: a tool to understand complexity". The
authors are grateful to M.R. Zirnbauer  for discussions and
hospitality extended to them at the initial stage of the present
research which took place at the Institute of Theoretical Physics,
University of Cologne, Germany.

\appendix

\section{Proof of Eq.~\eqref{eq:o2,1}}
\label{app:completeo21} Introducing the set of new variables  \be
a=\frac{1}{2}(p_1-p_2),\quad b=\frac{1}{2}(p_1+p_2), \quad
c=p_3\label{eq:abc}
\end{align}
and defining $t=\frac{r^2}{1-r^2}$, we can rewrite
Eq.~\eqref{eq:inapp} as \be\label{eq:app-int}
I_{HS}^{\mathrm{O}(2,1)}=\int_0^{\infty}\!\!\frac{dt}{\sqrt{1+t}}
&\int_0^\infty\!\! da\int_{-\infty}^\infty dbdc\ a[(b-c)^2-a^2]
\exp\bigg\{-a^2-b^2-\frac{c^2}{2}+i[x(b+c)+zb]\nn\\&+i\cos2 \theta
\ ta(x-z)+i(1+t)(x-z)(b-c)\bigg\}J_0(\sqrt{A^2+B^2})\, ,
\end{align}
where we have used Eq.~\eqref{eq:dP}. Next, we verify that \be
A^2+B^2&=w^2\bigg\{t^2(b-c)^2+2t(t+2)\cos2\theta\
a(b-c)+a^2[t^2\cos^22\theta+4(t+1)]
\bigg\}\nn\\
&=w^2\bigg\{[t(b-c)+a(t+2)\cos2\theta]^2+a^2[4(t+1)\sin^22\theta]
\bigg\}\nn\\&=C^2+D^2\, ,
\end{align}
where we defined \be
C=&w[t(b-c)+a(t+2)\cos2\theta]\nn\\D=&2wa\sqrt{t+1}\sin2\theta\, .
\end{align}
This allows us to write  (cf. (\ref{bessel11})) \be
\label{bessel1} J_0(\sqrt{A^2+B^2})=\frac{1}{\pi}\int_0^\pi d\phi
\exp\big(i\cos\phi C+i\sin\phi D\big)\, .
\end{align}
Using this representation of the Bessel function in
Eq.~\eqref{eq:app-int} and defining $y=1/\sqrt{1+t}$, we can
readily carry out integrals over $a$, $b$, $c$ and $\theta$, and
get \be\label{eq:final-o21}
I_{HS}^{\mathrm{O}(2,1)}=\exp[-x^2-\frac{z^2}{2}]F(w)\, ,
\end{align}
where we defined
\be F(w)=\int_1^\infty dy\int_0^\pi d\phi \exp\bigg\{-[w^2-(x-z)^2]y^2-[w\cos\phi(y^2-1)+(x-z)y^2]^2\bigg\}\, .
\end{align}
Note that although the above integral formally seems to depend on
both $w$ and $(x-z)$, we shall see below that it is a function of
$w$ only and is actually independent of the second combination.

To calculate $F(w)$ we find it convenient to apply first the
standard $\hs$ transformation and "linearise" the second term in
the exponent by introducing an auxiliary Gaussian integral: \be
F(w)=\int_1^\infty dy\int_0^\pi d\phi
\int_{-\infty}^{\infty}dh\exp\bigg\{&-[w^2-(x-z)^2]y^2-h^2\nn
\\ &+2ih\ [w\cos\phi(y^2-1)+(x-z)y^2]\bigg\}\, .
\end{align}
Integration over $\phi$ yields the Bessel function which can be
expanded in its Taylor series, and the Gaussian integral over $h$
can be performed. In this way we find \be F(w)={\mathrm{const}}
\sum_{n=0}^{\infty}w^{2n}\!\int_1^\infty\!\! dy\
e^{-(x-z)^2(y^4-y^2)}\bigg\{\left[\frac{1}{y^2}-2(1-2y^2)(x-z)^2\right]C_n+2(1-2y^2)
C_{n-1}\bigg\}\label{eq:fw}
\end{align}
where we defined for $n\ge0$ \be
C_n=\sum_{m=0}^{n}\frac{(-)^{n-m}}{(n-m)!}y^{2(n-m)}(y^2-1)^{2m}
\frac{(2m)!}{m!m!}\sum_{k=0}^m\frac{(-)^k}{4^kk!}\frac{[(x-z)y^2]^{2m-2k}}{(2m-2k)!}
\end{align}
and $C_n=0$ for $n<0$. In particular, the definition above implies
\be C_n\bigg\lvert_{y=1}=\frac{(-)^n}{n!}\, .
\end{align}
 $F(w)$ can be found as we are now able to perform the integrations on the right-hand
side of Eq.~\eqref{eq:fw} as  \be\label{eq:1n} &\int_1^\infty\!\!
dy\ e^{-(x-z)^2(y^4-y^2)}
\bigg\{[\frac{1}{y^2}-2(1-2y^2)(x-z)^2]C_n+2(1-2y^2)C_{n-1}\bigg\}\nn
\\=&-\frac{e^{-(x-z)^2(y^4-y^2)}}{y}
\big[C_n+y^2(y^2-1)\sum_{i=0}^{n-1} a_{n,i} C_{i}\big]
\bigg\lvert_{y=1}^\infty\nn\\=&\frac{(-)^n}{n!}\, .
\end{align}
Here $a_{n,i}$'s are coefficients satisfying the following
recursive relations \be a_{n,n-1}=\frac{2}{n},\ \ \ \mathrm{and}\
\ \ a_{n.i}=-\frac{1}{n}a_{n-1,i}, \ \ \ \ i=0,\dots,n-2,\ \ \
a_{1,0}=2.
\end{align}
In the last step of Eq.~\eqref{eq:1n} we used the fact $x-z > 0$.
We finally see that Eq.~\eqref{eq:1n} implies the desired Gaussian
expression \be F(w)={\mathrm{const }}\, e^{-w^2}
\end{align}
Finally, substituting $F(w)\propto e^{-w^2}$ back to
Eq.~\eqref{eq:final-o21} completes our proof of
Eq.~\eqref{eq:o2,1}.

\section{Calculation with the naive choice of the volume element $D\hat{P}$ for O$(2,1)$ case}
\label{app:check1}

In this appendix, we show the $\hs$ transformation for the O$(2,1)$ $\ps$ domain is invalid if
the volume element is chosen to be $D\hat{P}=|\Delta(\hat{P})|\prod_{i=1}^3dp_i$.\\

Starting from Eq.~\eqref{eq:sub}, we make a change of integration variables as in Eq.~\eqref{eq:abc}.
Then we write Eq.~\eqref{eq:sub} as
\be \mathcal{I}_{HS}^{\mathrm{O}(2,1)}=\int_{0}^{\infty}\frac{dt}{\sqrt{1+t}}
\int_{-\infty}^{\infty} \mathcal{D}\hat{P}\ e^{-a^2-b^2-\frac{c^2}{2}+i(2xb+zc)+it(x-z)(b-c)}
J_0\left[t(x-z)a\right]\, ,
\end{align}
where
\be D\hat{P}=2|a((b-c)^2-a^2)|\s da\s db\s dc\, .
\end{align}
We rewrite the above integral as \be
\mathcal{I}_{HS}^{\mathrm{O}(2,1)}=\mathcal{I}_{HS,1}^{\mathrm{O}(2,1)}+
\mathcal{I}_{HS,2}^{\mathrm{O}(2,1)}\, ,
\end{align}
where we defined
\be \mathcal{I}_{HS,1}^{\mathrm{O}(2,1)}=
\int_{0}^{\infty}\frac{dt}{\sqrt{1+t}}\int_{-\infty}^{\infty}&
2|a|((b-c)^2-a^2)\s da\s db\s dc\
e^{-a^2-b^2-\frac{c^2}{2}+i(2xb+zc)+it(x-z)(b-c)}\nn\\
&J_0\left[t(x-z)a\right]
\end{align}
and
\be\mathcal{I}_{HS,2}^{\mathrm{O}(2,1)}=
\int_{0}^{\infty}\!\!\frac{dt}{\sqrt{1+t}}\int_{0}^{|b-c|}&\!\!\!
4|a|(a^2-(b-c)^2)\s da\int_{-\infty}^{\infty}\!\! dbdc\
e^{-a^2-b^2-\frac{c^2}{2}+i(2xb+zc)+it(x-z)(b-c)}\nn\\
&J_0\left[t(x-z)a\right].
\end{align}
Let us stress that it is the contribution
$\mathcal{I}_{HS,2}^{\mathrm{O}(2,1)}$ which encapsulates the
difference between the definition
$D\hat{P}=|\Delta(\hat{P})|\prod_{i=1}^3dp_i$ which is positive
definite and Eq.~\eqref{eq:dP} which is sign indefinite. Such a
term has cancelled out when we the volume element was chosen to be
Eq.~\eqref{eq:dP}. The first contribution
$\mathcal{I}_{HS,1}^{\mathrm{O}(2,1)}$ is nothing else but the
$I_{HS}^{\mathrm{O}(2,1)}$ calculated in the Section 2, and we
proved it is in the Gaussian form
\be\mathcal{I}_{HS}^{\mathrm{O}(2,1)}=
const.\exp\bigg[-\frac{1}{2}(2x^2+z^2)
\bigg]+\mathcal{I}_{HS,2}^{\mathrm{O}(2,1)}\s.
\end{align}

In the remaining part of this appendix we will demonstrate that
the first contribution $\mathcal{I}_{HS,2}^{\mathrm{O}(2,1)}$ is
not in a Gaussian form, thus invalidating
the Hubbard-Stratonovich transformation.\\

Define $m=b+c$, $n=b-c$ and integrate over $m$. We get\be \mathcal{I}_{HS,2}^{\mathrm{O}(2,1)}=&
\frac{2\sqrt{6\pi}}{3}\exp\big[-\frac{1}{6}(2x+z)^2\big]\int_{-\infty}^{\infty}dn\
n^4\exp\bigg\{-(\frac{1}{3}+a^2)n^2+in(x-z)(\frac{2}{3}+t)\bigg\}
\nn\\
&\int_0^1da\int_0^\infty\frac{dt}{\sqrt{1+t}}a(a^2-1)J_0\bigg[t(x-z)na\bigg]\s.\end{align}
It is clear that $\mathcal{I}_{HS,2}^{\mathrm{O}(2,1)}$ will be in
the desired Gaussian form if integral part of the above formula is
$\propto \exp(-(x-z)^2/3)$. To check this, it is sufficient to
consider a special case $x\to z$, i.e. $|x-z|\ll 1$. In this
limit, we can approximate the integral by setting the argument of
the Bessel function in the integrand to zero. This gives \be
\mathcal{I}_{HS,2}^{\mathrm{O}(2,1)} \propto
e^{-\frac{(2x+z)^2}{6}}&\int_0^\infty\frac{dt}{\sqrt{1+t}}\int_{-\infty}^{\infty}dn\
n^4\exp\bigg\{-\frac{1}{3}n^2+in(x-z)(\frac{2}{3}+t)\bigg\}\nn\\
&\int_0^1da a(a^2-1)\exp(-n^2a^2)\, .
\end{align} The integral over $a$ is simply \be
\int_0^1da\s
a(1-a^2)\exp(-n^2a^2)=\frac{1}{2n^4}[\exp(-n^2)+n^2-1].\end{align}
Carrying out the standard Gaussian integrals over $n$, we get \be
\mathcal{I}_{HS,2}^{\mathrm{O}(2,1)} \propto
e^{-\frac{(2x+z)^2}{6}}\int_0^\infty\frac{dt}{\sqrt{1+t}}
\bigg[\frac{1}{2}\big[(x-z)^2(3t+2)^2-2\big]& e^{-\frac{1}{12}(x-z)^2(3t+2)^2}\nn\\
&-e^{-\frac{1}{48}(x-z)^2(3t+2)^2}\bigg]\, .\end{align} The integral
over $t$ is divergent if $x-z=0$, as expected, and in the limit
$|x-z|\ll 1$ it is a well-defined expression dominated by $t\sim
(x-z)^{-1} \gg 1$ so that $\mathcal{I}_{HS,2}^{\mathrm{
O}(2,1)}\sim (x-z)^{-1/2}$. Such a pre-exponential factor clearly
precludes the expression to be in the desired Gaussian form.

\section{Calculations for the standard case O$(3)$}
\label{app:o3}

In this appendix, we repeat calculations similar to those in
section 2 and App.~\ref{app:check1}, but this time for the compact
case of O$(3)$ group. Although the $\hs$ transformation for O$(3)$
symmetry is trivially valid in the original formulation, it is
instructive to have a comparison between O$(3)$ and O$(2,1)$ in
the polar representation, as it helps to
 understand peculiarities of the non-compact case.\\

First, we consider an integral similar to Eq.~\eqref{eq:break},
with integration of $\hat{T}$ going this time over $\mathrm{O}(3)$
instead of O$(2,1)$. We consider only the simplest case setting
$\hat{A}=\diag(x,x,z)$ and deal with the following integral \be
\label{eq:o3} I_{HS}^{\mathrm{O}(3)}=\int \mathcal{D}\hat{R}\
e^{-\frac{1}{2}\tr\hat{R}^2 -i\tr
\hat{R}\hat{A}}=\int_{-\infty}^{\infty} \mathcal{D}\hat{P}
e^{-\frac{1}{2}\sum\limits_{i=1}^{3}p_i^2}\int_{G/H}d\mu(\hat{S})\
e^{-i\tr\hat{S}^{-1}\hat{P}\hat{S}\hat{A}}\, ,
\end{align}
where $G=\mathrm{O}(3)$ and $H=\mathrm{O}(2)\times\mathrm{O}(1)$.
Elements of this compact coset is parametrised as \be
\label{eq:s2} s=g_H=\left(\begin{array}{cc}
(1+ZZ^T)^{-\frac{1}{2}} & Z(1+Z^TZ)^{-\frac{1}{2}}\\
Z^T(1+ZZ^T)^{-\frac{1}{2}} & (1+Z^TZ)^{-\frac{1}{2}}
\end{array}\right),
\end{align}
where we introduced the $2\times1$ real matrix $Z$ as the
convenient coordinate on $G/H$, with
\be Z=\left(\begin{array}{c} z_1 \\
z_2\end{array}\right)\,,\quad \mbox{with}\quad \ {\mathrm{z_1\
and\ z_2\ arbitrary\ real.}}
\end{align}
Similar to the non-compact case, $s^{-1}(Z,Z^T)=s(-Z,-Z^T)$. The
invariant measure $d\mu(\hat{S})$ in the projective coordinates is
given by \be
d\mu(\hat{S})=\frac{dZdZ^T}{(1+Z^TZ)^{\frac{3}{2}}}\,,\quad\mbox{where}\quad
dZdZ^T=dz_1dz_2\,.
\end{align}
The integration over the coset is now straightforward and
calculations are done
parallel to those in section 2. After some algebra and a few
changes of variables, we get \be
I_{HS}^{\mathrm{O}(3)}=\int_{0}^{1}\frac{dt}{\sqrt{1-t}}\int_{-\infty}^{\infty}
\mathcal{D}\hat{P}\ &\exp\bigg\{-\frac{1}{2}\sum\limits_{i=1}^{3}p_i^2+i(x(p_1+p_2)+zp_3)\nn\\
&+\frac{it}{2}(z-x)(p_1+p_2-2p_3)\bigg\} J_0\left[\frac{t}{2}(x-z)(p_1-p_2)\right]\, .
\label{eq:sub2}
\end{align}
The difference between Eq.~\eqref{eq:sub2} and Eq.~\eqref{eq:sub}
is due to the difference between
compact and non-compact integration manifolds.\\

A crucial difference in the O$(3)$ case is that the volume
elements $\mathcal{D}\hat{P}$ in the above formula is
$\mathcal{D}\hat{P}=|\Delta(\hat{P})|\prod_{i=1}^3dp_i$, instead
of Eq.~\eqref{eq:dP}. We have seen in App.~\ref{app:check1} that
this choice of $\mathcal{D}\hat{P}$ when applied for O$(2,1)$
symmetry would yield a form which is not Gaussian. In the
remaining part of this appendix we show that in the case of
 O$(3)$ the result is in contrast Gaussian.\\

Define the same set of integration variables as Eq.~\eqref{eq:abc}
and use them in Eq.~\eqref{eq:sub2}. We have \be
\mathcal{I}_{HS}^{\mathrm{O}(3)}=
\mathcal{I}_{HS,1}^{\mathrm{O}(3)}+\mathcal{I}_{HS,2}^{\mathrm{O}(3)}\, ,
\end{align}
where we defined
\be
\mathcal{I}_{HS,1}^{\mathrm{O}(3)}=
\int_{0}^{1}\frac{dt}{\sqrt{1-t}}\int_{-\infty}^{\infty}& 2|a|
((b-c)^2-a^2)\s da\s db\s dc\ e^{-a^2-b^2-\frac{c^2}{2}+i(2xa+zc)-it(x-z)(a-c)}\nn\\
 &J_0\left[t(x-z)a\right]
 \end{align}
and
\be
\mathcal{I}_{HS,2}^{\mathrm{O}(3)}=\int_{0}^{1}\!\!
\frac{dt}{\sqrt{1-t}}\int_{0}^{|b-c|}&\!\!\! 4|a|(a^2-(b-c)^2)\s
da\int_{-\infty}^{\infty}\!\! dbdc\ e^{-a^2-b^2-\frac{c^2}{2}+i(2xa+zc)-it(x-z)(a-c)}\nn
\\ &J_0\left[t(x-z)a\right]\s.
\end{align}
Note again that $\mathcal{I}_{HS,1}^{\mathrm{O}(3)}$ corresponds to the definition Eq.~\eqref{eq:dP} and
 $\mathcal{I}_{HS,2}^{\mathrm{O}(3)}$ emerges only because the volume element is positive
 definite in the current case.\\

First, we deal with $\mathcal{I}_{HS,1}^{\mathrm{O}(3)}$. Carrying
out simple Gaussian integrations over $a$, $b$ and $c$ we find \be
\mathcal{I}_{HS,1}^{\mathrm{O}(3)}=
\frac{\sqrt{2}\pi}{32}\mathcal{F}_1(x,z)
e^{-\frac{1}{2}(2x^2+z^2)}\label{eq:pre-gauss-2}\, ,
\end{align}
where
\be\mathcal{F}_1(x,z)=\int_{0}^{1}\frac{dt}{\sqrt{1-t}}
\exp\left(-\frac{1}{2}(t^2+t)(x-z)^2\right)\left[1-(x-z)^2(2t^2+3t+1)\right]\, .
\end{align}
Using $a=(x-z)^2$ and $y=\sqrt{1-t}$ we immediately see that
\be\mathcal{F}_1(a)&=\int_{0}^{1}\frac{dt}{\sqrt{1-t}}
\exp\left(-\frac{1}{2}(t^2+t)(x-z)^2\right)
\left[1-(x-z)^2(2t^2-3t+1)\right]\nn\\
&=\int_{0}^{1}dy\ \exp\left(-\frac{a}{2}(y^4-y^2)\right)
\left[1-a(2y^4-y^2)\right]\nn\\&=1-\lim_{y\to
0}y\exp(-\frac{ay^2(y^2-1)}{2})=1\, .
\end{align}
Here, the integral over $y$ is the same as the one in Eq.~\eqref{eq:o21} with different upper and
lower limits, but the result is the same. This completes our proof that $\mathcal{I}_{HS,1}^{\mathrm{O}(3)}$
indeed in the Gaussian form. We also note there is certain kind of duality between $\mathcal{I}_{HS,1}^{\mathrm{O}(3)}$
and $\mathcal{I}_{HS}^{\mathrm{O}(2,1)}$.\\

As we know already, the integral
$\mathcal{I}_{HS}^{\mathrm{O}(3)}$ is of the Gaussian form
$\propto\exp[-x^2-z^2/2]$, and we have just shown that the same
holds for $\mathcal{I}_{HS,1}^{\mathrm{O}(3)}$, the second
terms $\mathcal{I}_{HS,2}^{\mathrm{O}(3)}$ can then only be either
0 or the same Gaussian form as
$\mathcal{I}_{HS,1}^{\mathrm{O}(3)}$. To see which is the case, it
is sufficient to consider the same limit $x\to z$ as we did in
App.~\ref{app:check1}. In the limit $|x-z|\ll 1$, we find \be
\mathcal{I}_{HS,2}^{\mathrm{O}(3)}=\int_{0}^{1}\!\!\frac{dt}{\sqrt{1-t}}
\int_{0}^{|b-c|}&\!\!\! 4|a|(a^2-(b-c)^2)\s da
\int_{-\infty}^{\infty}\!\! dbdc\
e^{-a^2-b^2-\frac{c^2}{2}+i(2xa+zc)}\,.
\end{align}
One can perform all the integrations in this formula explicitly,
and show that \be\mathcal{I}_{HS,2}^{\mathrm{O}(3)}\propto
\exp\big\{-\frac{1}{2}(2x^2+z^2)\big\}
\end{align}
as expected.

\section{Jacobian of the transformation from $Z$ to polar coordinates}
\label{app:jacobi}
Write the polar coordinates decomposition in Eq.~\eqref{eq:polar} as $Z=O_1\Lambda O_2$, where
$\Lambda=\left(\begin{array}{cc}r&0\\0&s\end{array}\right)$. We have
\be \left\{ \begin{array}{l}
dZ\ \ =dO_1\Lambda O_2+O_1d\Lambda O_2+O_1\Lambda dO_2\\
dZ^T=dO_2^T\Lambda O_1+O_2^Td\Lambda O_1+O_2^T\Lambda dO_1\end{array}
\s.\right.
\end{align}
Following the standard way of derivation, see e.g.\cite{hua}, we
have \be d^2S =\tr\ dZdZ^T=\tr\ &\bigg\{ O_1^TdO_1\Lambda
O_2dO_2^T\Lambda+ \Lambda d\Lambda
O_2dO_2^T+\Lambda^2dO_2dO_2^T\nn\\& +O_1^TdO_1\Lambda
d\Lambda+d^2\Lambda+ \Lambda d\Lambda dO_2O_2^T\nn\\&
+\Lambda^2dO_1^TdO_1+\Lambda d\Lambda dO_1^TO_1+ \Lambda
dO_2O_2^T\Lambda dO_1^TO_1\bigg\}.
\end{align}

Next we define
\be\left\{ \begin{array}{l}
\delta O_1 =O_1^TdO_1\\
\delta O_2=dO_2O_2^T\end{array}\right.\, .
\end{align}
Recalling that $\delta O_1$ and $\delta O_2$ are skew-symmetric
matrices, which can be written as
\be \delta O_1=\left(\begin{array}{cc}0&\delta O_{1,12}\\-\delta O_{1,12}&0 \end{array}\right)
, \hspace*{5mm} \delta O_2=\left(\begin{array}{cc}0&\delta O_{2,12}\\-\delta O_{2,12}&0 \end{array}\right)\, .\end{align}
We find \be d^2S &=\tr \bigg\{ d^2\Lambda-\Lambda^2\delta
O_1\delta O_1-\Lambda^2 \delta O_2\delta O_2-2\Lambda \delta
O_1\Lambda \delta O_2\bigg\}\nn\\
& =(\delta O_{1,12}, \delta O_{2,12}, dr, ds)\left(\begin{array}{cccc}r^2+s^2 & 2rs&&\\2rs&r^2+s^2&&\\&&1&\\&&&1 \end{array}
\right)\left(\begin{array}{c}\delta O_{1,12}\\
\delta O_{2,12}\\ dr\\ ds \end{array} \right)\nn\\&=dx^i g_{ij}
dx^j\, .\end{align} In the last step the summation over repeated
indices is assumed. Jacobian is then given by \be
\mathrm{Jacobian}=\sqrt{\det g}=|r^2-s^2|\s.
\end{align}

\section{Calculation with the alternative volume element $D\hat{P}$ for O$(2,2)$ case}
\label{app:check2}

In this appendix, we calculate Eq.~\eqref{eq:o22} with the volume
element $D\hat{P}=|\Delta[\hat{P}]|\prod_{i=1}^4dp_i$ used instead
of Eq.~\eqref{eq:dP2}. We show that by this choice the final
result is not in the Gaussian form,
hence the corresponding $\hs$ transformation can not be valid.\\

First we redefine the integration variables
\be\left\{\begin{array}{l}a=
\frac{1}{2}(p_1+p_2)\\b=\hspace*{0.5mm}\frac{1}{2}(p_1-p_2)
\\c=\hspace*{0.5mm}\frac{1}{2}(p_3+p_4)\\ d=\frac{1}{2}(p_3-p_4)\s.\end{array}\right.
\end{align}
Then we have
\be |\Delta[\hat{P}]|=4|bd|\!\cdot\!|[(a-c+d)^2-b^2][(a-c-d)^2-b^2]|.
\end{align}
Use Eq.~\eqref{eq:coset2} and the $D\hat{P}$ defined above to find
\be \mathcal{I}_{HS}^{\mathrm{O}(2,2)}=\int_1^\infty
&\frac{|u-v|dudv}{\sqrt{u(u-1)}
\sqrt{v(v-1)}}\int D{\hat{P}} \ e^{-a^2-b^2-c^2-d^2+i(x-z)(a-c)(u+v)}\nn\\
&J_0\left[b(x-z)(u-v)\right]J_0\left[d(x-z)(u-v)\right].
\end{align}
As in App.~\ref{app:check1} we can split
$\mathcal{I}_{HS}^{\mathrm{O}(2,2)}$ into two parts, \be
\mathcal{I}_{HS}^{\mathrm{O}(2,2)}=\mathcal{I}_{HS,1}^{\mathrm{O}(2,2)}+
\mathcal{I}_{HS,2}^{\mathrm{O}(2,2)}\s.
\end{align}
Here, the contribution $\mathcal{I}_{HS,1}^{\mathrm{O}(2,2)}$ is
precisely the $I_{HS}^{\mathrm{O}(2,2)}$ we calculated in section
3, and we know it is in a Gaussian form. It is the second
contribution, $\mathcal{I}_{HS,2}^{\mathrm{O}(2,2)}$, which arises
from the difference between the two definitions of $D\hat{P}$, and
it is given by \be \mathcal{I}_{HS,2}^{\mathrm{O}(2,2)}\propto&
\int_{-\infty}^{\infty}dadc\int_0^\infty dd\int_{||a-c|-d|}^{|a-c|+d}db\ bd[(a-c+d)^2-b^2][(a-c-d)^2-b^2]\nn\\
&\int_1^\infty \frac{|u-v|dudv}{\sqrt{u(u-1)}\sqrt{v(v-1)}}
\ e^{-a^2-b^2-c^2-d^2+2ixc+2iza+i(x-z)(a-c)(u+v)}\nn\\
& \hspace{8mm}J_0\left[b(x-z)(u-v)\right]J_0\left[d(x-z)(u-v)\right]\s.
\end{align}In the remaining part of this section we demonstrate that
$\mathcal{I}_{HS,2}^{\mathrm{O}(2,2)}$ is not in the Gaussian
form, thus $\mathcal{I}_{HS}^{\mathrm{O}(2,2)}$ is not either.
Which means that $\hs$ transformation
fails with this different choice of $D\hat{P}$.\\

First, we define $m=a+c$ and $n=a-c$. It is clear that the
integral over $m$ is decoupled from other integrations and can be
easily performed. Again, it is sufficient to consider the limit
$|x-z|\ll 1 $. For the same reason as in App.~\ref{app:check1}, we
set the two Bessel terms to be 1 in this limit. We then have \be
\mathcal{I}_{HS,2}^{\mathrm{O}(2,2)}\propto&\exp\{-\frac{1}{2}(x+z)^2\}\int_0^\infty\!
dn \int_0^\infty \!dd\int_{|n-d|}^{n+d}\!db\
bd[(n+d)^2-b^2][(n-d)^2-b^2]\nn\\ &\int_1^\infty\!
\frac{|u-v|dudv}{\sqrt{u(u-1)}\sqrt{v(v-1)}}\exp\left\{-n^2-b^2-d^2\right\}\cos\left[n(x-z)(u+v-1)
\right]\, .
\end{align}The integral part of the above formula needs to be $\propto \exp\{-\frac{1}{2}(x-z)^2\}$ in order to make
$\mathcal{I}_{HS,2}^{\mathrm{O}(2,2)}$ be Gaussian. The remaining
calculations are lengthy but direct. We perform Gaussian type
integrals over $b$, $d$ and $n$ then define new integration
variables $X=u+v-1$ and $Y=u-v$. After integrating over $Y$ we get
\be \mathcal{I}_{HS,2}^{\mathrm{O}(2,2)}\propto\int_0^\infty dX
\ln(2X+1)\big\{8-2a^2X^2+\sqrt{\pi}e^{-\frac{a^2X^2}{4}}
aX(a^2X^2-6){\mathrm{Erfi}}[\frac{aX}{2}]\big\}\, ,\end{align} where
we defined $a=x-z$, and Erfi stands for the error function of
imaginary argument. Integrating by parts we bring the above
integral to the form \be
\mathcal{I}_{HS,2}^{\mathrm{O}(2,2)}\propto\int_0^\infty dX\
\frac{1}{2X+1}
\bigg(\frac{aX}{4}-\frac{\sqrt{\pi}}{8}e^{-\frac{a^2X^2}{4}}(a^2X^2-2){\mathrm{Erfi}}
[\frac{aX}{2}]\bigg)\, .\end{align} Again, in the limit $|x-z|\ll 1$
the integral over $X$ is dominated by the region
$X\sim(x-z)^{-1}\gg 1$. Changing variable $aX\to X$ and expanding
in terms of $|x-z|$, we find to the lowest order,
$\mathcal{I}_{HS,2}^{\mathrm{O}(2,2)}=c_0 -|x-z|c_1+O((x-z)^2)$,
with $c_0$ and $c_1$ being some constants. In this way one finds
that $\mathcal{I}_{HS,2}^{\mathrm{O}(2,2)}$ is a function of
$(x-z)$, but is clearly not in the Gaussian form. So we conclude
that $\mathcal{I}_{HS}^{\mathrm{O}(2,2)}$ is not Gaussian.


\begin{thebibliography}{99}
\bibitem{efe1} K.B. Efetov, {\em Supersymmetry in Disorder and Chaos},
(Cambridge University Press, 1997).
\bibitem{mirlin} A.D. Mirlin, Phys. Rep. {\bf 326} (2000) 259.
\bibitem{weg} F.Wegner, Zeitsch. Physik {\bf36} (1979) 207.
\bibitem{sw} L. Sch\"afer and F. Wegner, Z. Phys. B {\bf38} (1980) 113.
\bibitem{ps} A.M.M. Pruisken and L. Sch\"afer, Nucl. Phys. B {\bf200} (1982) 20.
\bibitem{efe} K.B. Efetov, Adv. Phys. {\bf32} (1983) 53.
\bibitem{vwz} J.J.M. Verbaarschot, H.A. Weidenm\"uller, M.R. Zirnbauer, Phys. Rep. {\bf129} (1985) 367.
\bibitem{fs} Y.V.Fyodorov and H.-J. Sommers, J. Math. Phys. {\bf38} (1997) 1918.
\bibitem{vw} J.J.M. Verbaarschot and T. Wettig, Annu. Rev. Nucl. Part. Sci. {\bf50} (2000) 343.
\bibitem{mehta} M-L. Mehta {\em Random Matrices}, 3rd ed, (Elsevier, 2004)
\bibitem{zirn-rev1}M.R. Zirnbauer, "The supersymmetry method of random matrix theory",
Encyclopedia of Mathematical Physics, vol. 5, p. 151-160,
Eslevier, 2006 [e-preprint arXiv:math-ph/0404057].
\bibitem{zirn-sss} M.R. Zirnbauer, J. Math. Phys. {\bf{37}} (1996) 4986.
\bibitem{yf1} Y.V. Fyodorov, J.Phys: Cond. Mat. {\bf 17} (2005) S1915.
\bibitem{zirn-rev2}M.R. Zirnbauer, "Symmetry classes in random matrix theory",
 e-preprint arXiv:math-ph/0404058.
\bibitem{fiz} Y.V. Fyodorov  Nucl. Phys. B {\bf 621} (2002) 643.
\bibitem{rigor} M.R. Zirnbauer, Y. Wei and Y.V. Fyodorov, under preparation.
\bibitem{iz} C.Itzykson and J.B. Zuber, J. Math. Phys. {\bf21} (1980) 411; Harish-Chandra,
Proc. Nat. Acad. Sci. {\bf42} (1956) 252.
\bibitem{fes} Y.V. Fyodorov and E. Strahov, Nucl. Phys. B {\bf630} (2002) 453.
\bibitem{GK} T. Guhr and H. Kohler, J. Math. Phys. {\bf43} (2002) 2707.
\bibitem{BH} E. Brezin and S. Hikami, Prog. of Theo. Phys. {\bf116} No.3 (2006) 441.
\bibitem{hua}L.K. Hua, {\em Harmonic analysis of functions of several complex numbers
in the classical domains}, (American Mathematical Society, Providence, 1963).
\bibitem{grad} I.S. Gradshteyn and I.M. Ryzhik, {\em Table of integrals, series, and products}, (Academic Press, 2000).
\end{thebibliography}
\end{document}